\documentclass[aps,pra,twocolumn,superscriptaddress]{revtex4-2}

\usepackage{amsmath,amssymb,bm,mathrsfs,amsfonts}
\usepackage{tikz}
\usepackage{pgfplots}
\usepackage{graphicx}
\usepackage{siunitx}
\usepackage{hyperref}
\usepackage{xcolor}

\usetikzlibrary{positioning, calc}
\pgfplotsset{compat=1.18}
\definecolor{editorred}{RGB}{180,0,0}

\begin{document}

\title{Floquet-tuned superfluid–checkerboard competition in dipolar bosons}

\author{Jin Yang}
\affiliation{Department of Physics, Anhui Normal University, Wuhu, Anhui 241000, China}
\author{Yaghmorassene Hebib}
\affiliation{Department of physical science, Butte College,  Oroville, California 95965, USA}
\author{Chao Zhang}
\email{chaozhang@ahnu.edu.cn}
\affiliation{Department of Physics, Anhui Normal University, Wuhu, Anhui 241000, China}


\begin{abstract}
We study hard-core dipolar bosons on a square lattice subject to a unidirectional periodic drive that Floquet-engineers anisotropic hopping.
Driving along one lattice direction provides a controlled way to suppress transverse tunneling, yielding a kinetically quasi-one-dimensional regime with strongly anisotropic transport within the leading-order high-frequency Floquet effective description. In this limit, the system does not reduce to decoupled chains, due to the long-range in-plane dipolar interaction remains isotropic and couples different chains.
Focusing on dipoles polarized perpendicular to the plane, for which the interaction is purely repulsive and isotropic, we use sign-problem-free worm-algorithm quantum Monte Carlo simulations to map the half-filling phase diagram versus kinetic anisotropy and dipolar coupling.
We find that increasing kinetic anisotropy systematically lowers the interaction strength required to stabilize checkerboard order, demonstrating that Floquet-induced suppression of transverse motion enhances density ordering.
Near the superfluid--checkerboard boundary, finite-size results reveal a narrow transition region where the stiffness drops rapidly while checkerboard correlations rise sharply; Its pronounced sharpening with system size is consistent with a weakly first-order transition rounded by finite-size effects.
Away from half filling, on the doped sides of the checkerboard plateau, we identify a narrow checkerboard-supersolid regime with simultaneously finite checkerboard correlations and superfluid stiffness, where the superfluid stiffness is anisotropic but the density pattern is isotropic.
\end{abstract}

\maketitle

\section{Introduction}
\label{sec:intro}

Periodic driving, or Floquet engineering, has emerged as a powerful and versatile tool for realizing exotic many-body states that are difficult to access in static systems~\cite{Eckardt2005,Lignier2007,Goldman2014,Eckardt2017}.
By subjecting a quantum system to a time-periodic modulation, one can dynamically renormalize its effective Hamiltonian parameters without altering the underlying microscopic interactions.
A paradigmatic example is the coherent control of tunneling amplitudes in shaken optical lattices, where the effective hopping can be tuned via Bessel-function renormalization and even strongly suppressed along selected spatial directions~\cite{Eckardt2005,Lignier2007}.
This capability makes Floquet engineering an efficient route for designing kinetic anisotropy and exploring dimensional crossover in a highly controlled manner.

Optical lattices loaded with ultracold atoms or polar molecules provide an ideal platform for implementing such Floquet protocols~\cite{Bloch2012,Lahaye2009,Goral2002,Moses2017}.
The exquisite control over lattice geometry, tunneling rates, and interaction strengths in these systems has enabled milestone experiments, including the observation of dipolar spin-exchange dynamics with lattice-confined polar molecules~\cite{Yan2013} and the realization of dipolar supersolidity and its collective dynamics in continuum quantum gases with strongly magnetic atoms~\cite{Tanzi2019PRL,Tanzi2019Nature,Guo2019Nature,Natale2019,Sohmen2021}.
These experimental advances sharpen a broader theoretical question: how do kinetic energy, dimensionality, and long-range repulsion conspire to produce superfluid, density-ordered, and supersolid phases in lattice realizations?
On the theoretical side, dipolar and more general extended-interaction boson models are known to host a rich competition between delocalization and crystalline ordering~\cite{Chanda_2025, Sinha_2025, Wessel2005, Goral2002,Yi2007,Capogrosso2010,Zhang2018PolarMolecules,Zhang2021TiltedDipolar,Hebib2023CavityDipolar,Hebib2024Thermocrystallization, Boninsegni2005}.
For isotropic square-lattice systems, a well-known knowledge has been established by various studies:
at half filling, the superfluid--checkerboard (SF--CB) transition is first order in the hard-core limit~\cite{Schmid2002,Capogrosso2010, Sengupta2005},
whereas away from commensurate filling, checkerboard supersolidity exists in finite doping windows~\cite{Ohgoe2012,Capogrosso2010,Yi2007}.
These results make the isotropic model a natural reference point for investigating how tunable kinetic anisotropy reshapes the location, character, and nearby supersolid tendencies of the SF--CB competition.
Although Floquet engineering has been successfully applied to the conventional Bose-Hubbard model~\cite{Eckardt2005,Lignier2007} and, more recently, to extended Hubbard models with nearest-neighbor interactions~\cite{PhysRevA.94.023615, Pieplow_2018}, its application to lattice models with genuine long-range dipolar interactions remains, to the best of our knowledge, unexplored.

In this work, we study hard-core dipolar bosons on a square lattice subject to a unidirectional periodic drive along one lattice direction (periodically shaking the lattice sites position along $y$ direction in this work). In the high-frequency regime, the drive renormalizes the hopping as $t_y^{\rm eff}=t\,J_0(K_y)$, where $J_0$ is the zeroth-order Bessel function of the first kind, and $K_y$ is the dimensionless shaking amplitude, while the transverse hopping $t_x=t$ remains unchanged.
This yields an effective anisotropic tunneling extended Bose-Hubbard model with dipolar interaction that interpolates continuously between an isotropic two-dimensional system and a kinetically quasi-one-dimensional regime of weakly coupled chains.
We focus on dipoles polarized perpendicular to the lattice plane, for which the in-plane dipolar interaction is purely repulsive and isotropic, $V_{ij}\propto 1/r_{ij}^3$.
This geometry is particularly transparent because any anisotropic many-body response can be traced directly to the Floquet-engineered kinetic anisotropy, rather than to intrinsic interaction anisotropy~\cite{Zhang_2015, Zhang2022, Yi2007, Zhang2021TiltedDipolar}.
Using sign-problem-free worm-algorithm quantum Monte Carlo simulations (WA-QMC), we map the half-filling phase diagram as a function of kinetic anisotropy and dipolar coupling, and extend the analysis to doped systems away from half filling.

Our main results are as follows.
First, we obtain an anisotropy--interaction phase diagram at half filling in the $(t_y^{\rm eff}/t_x, D/t)$ plane with $D$ is the dipolar interaction strength, and show that decreasing $t_y^{\rm eff}/t_x$ systematically lowers the dipolar interaction strength required to stabilize checkerboard order.
This demonstrates that Floquet-induced suppression of transverse motion strongly enhances interaction-driven crystallization even when the dipolar interaction itself remains isotropic.
Second, through finite-size scaling at fixed strong anisotropy, we find that the SF--CB transition occurs within a narrow interaction interval: for smaller system sizes the evolution appears relatively smooth, whereas for larger sizes it becomes markedly sharper, consistent with a weakly first-order transition rounded by finite-size effects.
Third, by analyzing the direction-resolved superfluid stiffnesses $\rho_{x}$ and $\rho_{y}$, we show that the approach to the SF--CB transition in the strongly anisotropic regime is accompanied by pronounced directional transport imbalance, with coherence along the weak-hopping direction suppressed much earlier than along the strong-hopping direction.
Finally, away from half filling we identify narrow checkerboard-supersolid regimes on both sides of the checkerboard plateau, where substantial diagonal order coexists with finite superfluid response.

The rest of the paper is organized as follows.
In Sec.~\ref{sec:model} we introduce the driven dipolar Bose--Hubbard model and derive the corresponding Floquet effective Hamiltonian.
Section~\ref{sec:method} summarizes the quantum Monte Carlo method and the observables used to characterize the phases.
Section~\ref{sec:results} presents the half-filling phase diagram, the finite-size evolution of the SF--CB transition, the associated transport anisotropy, and the checkerboard supersolid phase away from half filling.
Finally, Sec.~\ref{sec:conclusion} summarizes our findings and discusses their relevance for experiments on Floquet-engineered dipolar lattice bosons.

\section{Model and Floquet-engineered effective Hamiltonian}
\label{sec:model}

\begin{figure}[t]
    \centering
    \includegraphics[width=1.0\linewidth]{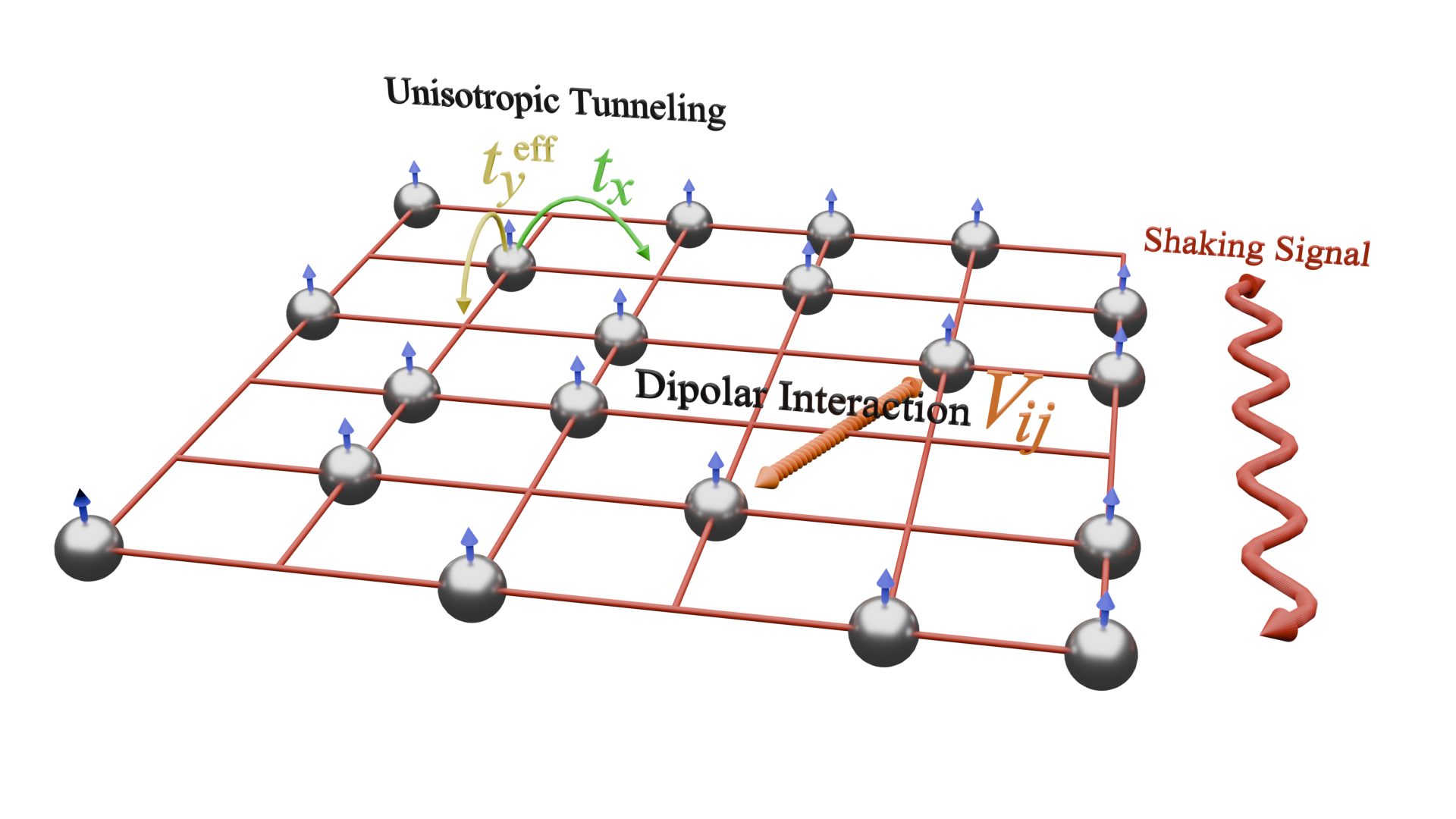}
   \caption{\textbf{Driven dipolar Bose--Hubbard setup and Floquet-induced hopping anisotropy.} Schematic of dipolar bosons in a shaken square optical lattice.
A periodic drive is applied along the $y$ direction, which in the lattice-gauge description is encoded by a time-dependent vector potential $A_y(t)=K_y\sin(\omega_m t+\phi)$ on the $y$ bonds, while the hopping along $x$ remains unchanged. Such scheme can be realized by periodically modulating the phase of the laser beam forming the standing wave along $y$ direction or even mechanical methods.
This realizes a controllable anisotropy between the two lattice directions.
The bosons carry dipole moments polarized perpendicular to the lattice plane. In this geometry, the in-plane dipolar interaction is isotropic and purely repulsive, $V_{ij}=D/r_{ij}^3$, where $r_{ij}$ is the distance between lattice sites $i$ and $j$ and $D$ sets the interaction strength.
    }
    \label{fig:model_setup}
\end{figure}

\subsection{Driven lattice and microscopic Hamiltonian}
We consider bosons on an $L_x\times L_y$ lattice with lattice spacing set to unity.
The system is subject to a unidirectional periodic drive (shaking) along the $y$ axis, which can be implemented by sinusoidally displacing the lattice position or equivalently introducing a time-periodic inertial force.
The model is illustrated in Fig.~\ref{fig:model_setup}.
A convenient lattice-frame description is
\begin{align}
\label{eq:H_t}
H(t)=&
-\!\!\sum_{\langle ij\rangle_x} t_x (b_i^\dagger b_j+\mathrm{H.c.})
-\!\!\sum_{\langle ij\rangle_y} t_y (b_i^\dagger b_j+\mathrm{H.c.}) \nonumber \\
&-\mu \sum_i n_i
+H_{\rm int},
\end{align}
where $b_i^\dagger$ creates a boson at site $i$, $n_i=b_i^\dagger b_i$, and $\langle ij\rangle_{x(y)}$ denotes nearest neighbors along $x(y)$. The tunneling amplitude is $t_x = t_y = t$, respectively. The interaction term includes on-site repulsion and dipolar long-ranged interaction,
\begin{equation}
\label{eq:Hint}
H_{\rm int}=\frac{U}{2}\sum_i n_i(n_i-1)
+\sum_{i< j} V_{ij}\, n_i n_j.
\end{equation}
We focus on polarized dipoles perpendicular to the 2D plane, for which the interaction is purely repulsive and isotropic,
\begin{equation}
\label{eq:Vdip}
V_{ij}=\frac{D}{r_{ij}^3},\qquad r_{ij}=|\mathbf r_i-\mathbf r_j|.
\end{equation}
Throughout the main text we concentrate on the hard-core limit $U\to\infty$ (restricting $n_i=0,1$).

Shaking the standing waves of the lattices along $y$ direction gives the atoms a periodic force $F_y(t)=F_0\cos(\omega_m t+\phi)$, where $\omega_m$ denotes the driving angular frequency (with period $T=2\pi/\omega_m$) and $\phi$ is the initial phase offset. In the lattice-gauge language, this corresponds to a time-dependent Peierls phase on the $y$-direction hopping.
Equivalently, in the accelerated frame one introduces a periodic vector potential $A_y(t)$ such that the $y$-bond hopping acquires a phase $e^{iA_y(t)}$.
For a sinusoidal drive, one may write
\begin{equation}
\label{eq:Ay}
A_y(t)=K_y\sin(\omega_{m} t+\phi),
\end{equation}
where $K_y$ is a dimensionless drive strength controlled by the shaking amplitude and frequency.
The $y$-direction kinetic term becomes
\begin{equation}
\label{eq:TyPeierls}
-\sum_{\langle ij\rangle_y} t_y \left( e^{iA_y(t)} b_i^\dagger b_j + \mathrm{H.c.}\right).
\end{equation}

\subsection{High-frequency Floquet expansion and effective anisotropic hopping}
In the high-frequency regime where $\hbar\omega_m$ is the dominant energy scale compared with the bare tunneling and interaction scales,
the driven system can be described by a static Floquet effective Hamiltonian.
To leading order in the inverse frequency, the effective Hamiltonian is obtained by averaging over a drive period $T=2\pi/\omega_m$.
Because the interaction term is diagonal in the occupation basis, the leading effect of the drive is the renormalization of hopping along the driven direction~\cite{Eckardt2005,Lignier2007}:
\begin{align}
\label{eq:teff_def}
t_y^{\rm eff}&
=t_y \left\langle e^{iA_y(t)}\right\rangle_T
=t_y \frac{1}{T}\int_0^T dt\, e^{iK_y\sin(\omega_m t+\phi)}  \nonumber \\
&=t_y J_0(K_y),
\end{align}
where $J_0$ is the zeroth order of the Bessel function of the first kind.
The hopping perpendicular to the drive remains unchanged at this order.
Thus the leading-order Floquet effective model reads
\begin{align}
\label{eq:Heff}
H_{\rm eff}=&
-\sum_{\langle ij\rangle_x} t_x (b_i^\dagger b_j+\mathrm{H.c.})
-\sum_{\langle ij\rangle_y} t_y^{\rm eff} (b_i^\dagger b_j+\mathrm{H.c.})
\nonumber \\
&-\mu \sum_i n_i
+\sum_{i< j}\frac{D}{r_{ij}^3}\, n_i n_j,
\end{align}
with the hard-core constraint $n_i\in\{0,1\}$ in the main calculations.
The ratio $t_y^{\rm eff}/t_x=J_0(K_y)\,(t_y/t_x) = J_0(K_y)$ can be tuned continuously by the drive strength $K_y$, enabling a controlled dimensional crossover: as $t_y^{\rm eff}\to 0$ the system enters a kinetically quasi-one-dimensional regime with strongly suppressed transverse tunneling, while the long-range dipolar interaction continues to couple different chains; by contrast, $t_y^{\rm eff}\simeq t_x$ corresponds to an isotropic 2D lattice.

In the present work, we restrict to the positive-hopping Floquet branch $0<K_y<K_0$ (with $K_0\approx2.4048$ the first zero of $J_0$), such that $t_y^{\rm eff}>0$ throughout.
This choice avoids crossing the Bessel zero and keeps the effective model in a sign-problem-free regime for our simulations. In the following we set $t=1$ as the unit of energy and study the phase structure of Eq.~\eqref{eq:Heff} by tuning the effective anisotropy ratio $t_y^{\rm eff}/t_x$ and dipolar coupling $D/t$.
At half filling, this allows us to resolve how Floquet-engineered kinetic anisotropy reshapes the competition between superfluid and checkerboard-solid order.

\section{Numerical method and observables}
\label{sec:method}

We consider square lattices with linear sizes $L=12,20,\dots,36$, inverse temperature $\beta=L$, and periodic boundary conditions. For the most anisotropic point $t_y^{\rm eff}/t_x=0.01$, we additionally use rectangular clusters to better accommodate anisotropic finite-size effects, following the aspect-ratio considerations discussed for anisotropic Bose--Hubbard models\cite{SchoenmeierKromer2014}.

We control the filling through the chemical potential $\mu$ in the grand-canonical ensemble.
The average density is
\begin{equation}
n=\frac{\langle N\rangle}{N_s}, \qquad N=\sum_i n_i, \qquad N_s=L_x L_y,
\end{equation}
and the compressibility is obtained from number fluctuations,
\begin{equation}
\label{eq:kappa}
\kappa=\frac{\beta}{N_s}\left(\langle N^2\rangle-\langle N\rangle^2\right).
\end{equation}
Incompressible phases are identified by plateaus in $n(\mu)$ together with $\kappa\simeq 0$, while compressible phases have finite $\kappa$.

Superfluidity is diagnosed from winding-number fluctuations.
Because the effective hopping is anisotropic, we define direction-resolved superfluid stiffnesses as
\begin{equation}
\label{eq:rhos_xy}
\rho_{x}=\frac{\langle W_x^2\rangle}{\beta\, t_x},\qquad
\rho_{y}=\frac{\langle W_y^2\rangle}{\beta\, t_y^{\rm eff}},
\end{equation}
where $W_x$ and $W_y$ are winding numbers along the $x$ and $y$ directions, respectively. The average superfluid stiffness $\rho_{\mathbf{ave}}=(\rho_{x}+\rho_{y})/2$. 
To quantify transport anisotropy induced by Floquet-engineered hopping anisotropy, we also use the dimensionless indicator
\begin{equation}
\label{eq:A_def}
R \equiv \frac{\rho_{x}-\rho_{y}}{\rho_{x}+\rho_{y}},
\end{equation}
for which $R\approx 0$ corresponds to nearly isotropic transport and $R\to 1$ indicates strongly suppressed superfluid response along $y$.

To detect long-range density-wave ordering on the square lattice, we compute the static structure factor at the checkerboard wave vector $\mathbf{k}=(\pi,\pi)$,
\begin{equation}
S(\mathbf{k})=\frac{1}{N_s}\sum_{\mathbf{r},\mathbf{r}'} e^{i \mathbf{k}\cdot(\mathbf{r}-\mathbf{r}')} \langle n_{\mathbf{r}} n_{\mathbf{r}'} \rangle ,
\end{equation}
A finite value of $S(\pi,\pi)$ in the thermodynamic limit indicates checkerboard charge-density-wave order.
Phase identification is based on the combined behavior of superfluid and density-order diagnostics.
We identify a superfluid region by finite stiffness and zero checkerboard correlations, and a checkerboard region by large $S(\pi,\pi)$ together with strong suppression of both $\rho_{x}$ and $\rho_{y}$ (and, when present, a half-filling plateau with $\kappa \simeq 0$).

\section{Results}
\label{sec:results}

In this section, we present the half-filling phase diagram, and analyze how kinetic anisotropy reshapes the competition between superfluid and checkerboard order.
We first map out the global phase diagram in the anisotropy--interaction plane $(t_y^{\rm eff}/t_x,D/t)$, which establishes the overall trend that reducing the Floquet-renormalized hopping $t_y^{\rm eff}/t_x$ lowers the interaction scale required to stabilize checkerboard order.

At finite system size, however, the SF--CB boundary is not represented by a sharp first-order transition line.
Instead, the numerical result shows the changeover into a narrow transition interval bounded by two nearby characteristic scales:
a lower scale at which the superfluid response begins to collapse rapidly, and an upper scale at which checkerboard order becomes clearly dominant.
For smaller system sizes, this interval appears relatively smooth and can mimic aspects of continuous critical behavior.
As the system size increases, however, the evolution becomes markedly sharper, indicating that this finite-width interval is more naturally interpreted as the finite-size-rounded manifestation of a weakly first-order transition.

We then move away from half filling and examine the doped sides of the checkerboard plateau.
There, the incompressible checkerboard solid gives way to regimes in which substantial diagonal checkerboard correlations coexist with a finite superfluid response.
This identifies supersolid windows adjacent to the half-filled checkerboard state, indicating that doping partially melts the commensurate density order without destroying it immediately, leading to the checkerboard supersolid phase emerging in the vicinity of $n=0.5$ upon doping.

\begin{figure}[t]
    \centering
    \includegraphics[width=\linewidth,trim=0.0cm 2.7cm 0.0cm 3.2cm, clip]{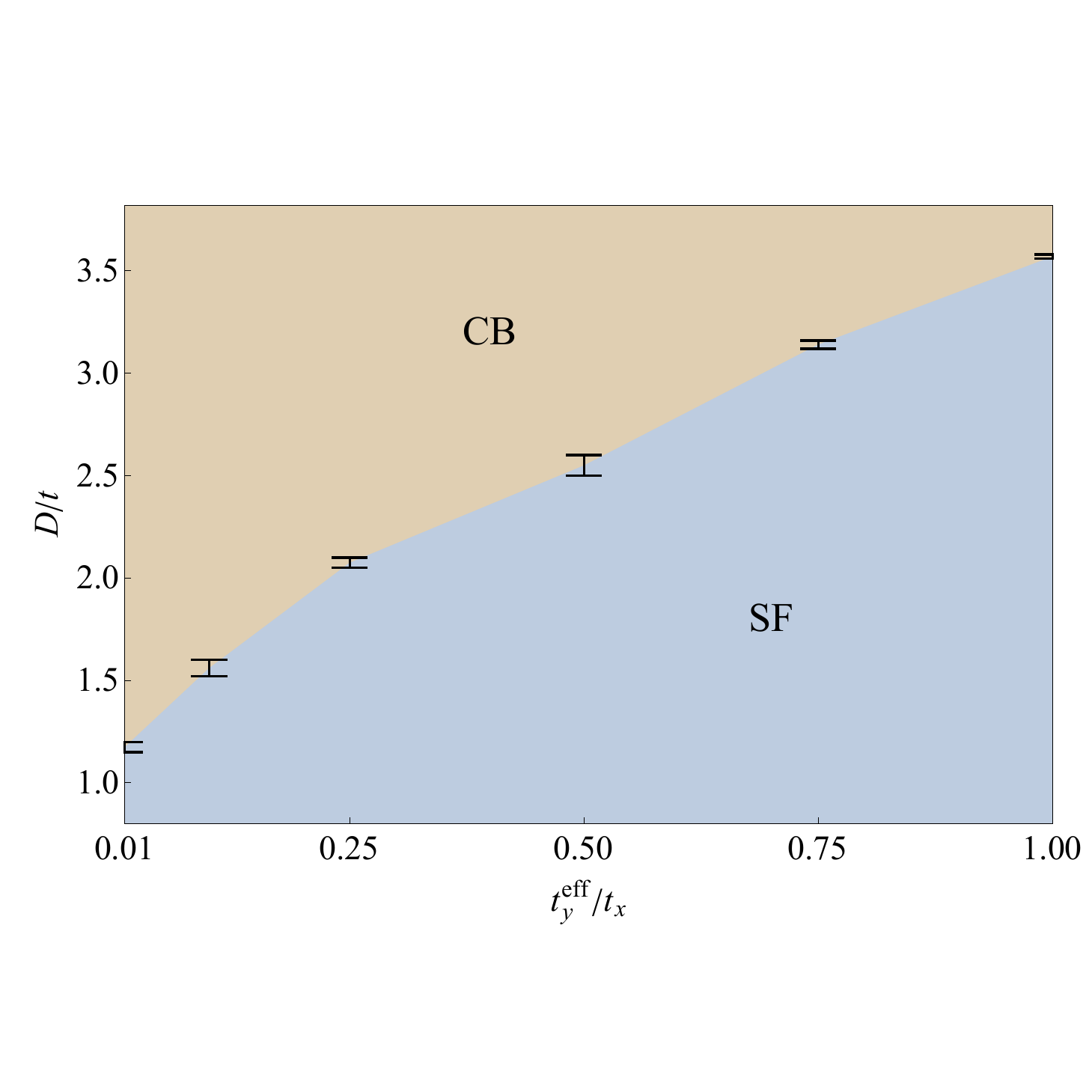}
    \caption{\textbf{Half-filling anisotropy--interaction phase diagram and transition-point uncertainty in the Floquet-engineered dipolar Bose--Hubbard model.}
    Phase diagram at fixed half filling $n=1/2$ in the plane of hopping anisotropy $t_y^{\rm eff}/t_x$ and dipolar interaction strength $D/t$ (with $t=1$).
    The lower blue region denotes the superfluid (SF) regime, and the upper tan region denotes the checkerboard solid (CB) regime.
    Symbols mark the estimated SF--CB transition points, while the vertical error bars denote the finite-size uncertainty window extracted from the onset of rapid stiffness collapse and the onset of robust checkerboard ordering. The lines connecting the points are guides to the eye.
    The monotonic upward shift of these transition points with increasing $t_y^{\rm eff}/t_x$ shows that stronger kinetic anisotropy (smaller $t_y^{\rm eff}/t_x$) stabilizes checkerboard order at lower dipolar coupling.
    }
    \label{fig:phase_diagram_main}
\end{figure}

\subsection{Half-filling phase diagram and weakly first-order SF--CB transition}
\label{subsec:results_phase_diagram}

We first establish the global half-filling phase diagram of the Floquet-engineered dipolar Bose--Hubbard model.
Figure~\ref{fig:phase_diagram_main} summarizes the phase diagram in the $(t_y^{\rm eff}/t_x,D/t)$ plane, where $t_x=t=1$ sets the energy scale and $t_y^{\rm eff}>0$ is the Floquet-renormalized hopping along the driven direction.
The phase identification is based on the combined behavior of the direction-resolved superfluid stiffnesses $\rho_{x}$ and $\rho_{y}$, together with the checkerboard structure factor $S(\pi,\pi)$.

The dominant trend is clear and monotonic:
as the hopping anisotropy becomes stronger (i.e., as $t_y^{\rm eff}/t_x$ decreases), the dipolar interaction scale required to destabilize the superfluid and favor checkerboard ordering is reduced.
Physically, reducing $t_y^{\rm eff}/t_x$ narrows the kinetic bandwidth and suppresses transverse delocalization, thereby increasing the relative importance of the dipolar repulsion and making checkerboard localization easier to stabilize.

Fig.~\ref{fig:phase_diagram_main} indicates the SF--CB transition is most naturally interpreted as weakly first order. In the phase diagram, this is represented by a set of estimated transition points with vertical error bars, rather than by an extended intermediate phase. Operationally, the central symbol marks the best estimate of the transition location, while the upper and lower ends of the error bar span the interval between the onset of rapid stiffness collapse and the onset of robust checkerboard ordering. This uncertainty range reflects the finite-size rounding of a weakly first-order transition in QMC data on finite lattices.

This representation is more faithful to the numerical results than a single sharp boundary line.
Near the transition, one typically observes a narrow interval in which the superfluid response is already strongly suppressed---highly anisotropic---while $S(\pi,\pi)$ rises rapidly but has not yet fully reached its large-$D/t$ behavior.
Rather than indicating a continuous second-order phase transition, this behavior is more naturally understood as the finite-size manifestation of phase competition across a weakly first-order SF--CB transition. The error bars in Fig.~\ref{fig:phase_diagram_main} therefore provide a compact summary of the finite-size uncertainty in locating the weakly first-order SF--CB boundary across the full anisotropy range.

In the following, we focus on the representative strongly anisotropic cut $t_y^{\rm eff}/t_x=0.1$, which exhibits a pronounced directional imbalance in transport and the clearest smooth growth of checkerboard order, and use the combined evolution of $\rho_{x}$, $\rho_{y}$, $R$, and $S(\pi,\pi)$ to characterize the weakly first-order SF--CB transition.


\begin{figure}[h]
    \centering
    \includegraphics[width=1.9\linewidth,trim=9.5cm 0.1cm 1cm 0.2cm, clip]
    {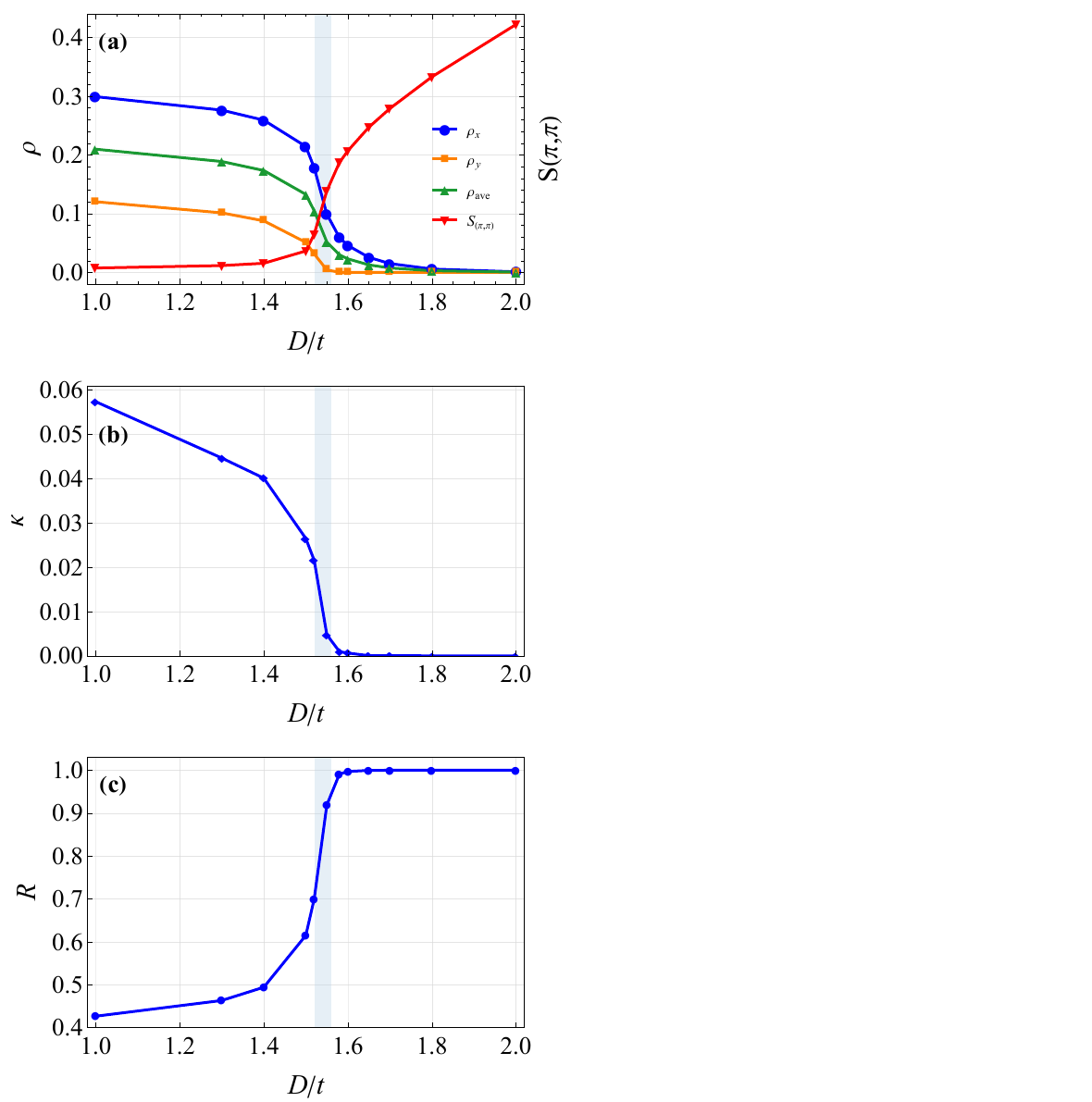}
    \caption{\textbf{Superfluid response, compressibility, and checkerboard correlations versus dipolar interaction at strong kinetic anisotropy ($t_y^{\rm eff}/t_x=0.1$, $L=20$).}
    (a) Direction-resolved superfluid stiffnesses $\rho_{x}$ and $\rho_{y}$, their average $\rho_{\mathbf{ave}}=(\rho_{x}+\rho_{y})/2$, and the checkerboard structure factor $S(\pi,\pi)$ as functions of dipolar interaction strength $D/t$ at half filling. (b) Compressibility $\kappa$ as a function of $D/t$. (c) Stiffness anisotropy $R=(\rho_{x}-\rho_{y})/(\rho_{x}+\rho_{y})$.
    Shaded bands mark the finite-size transition interval.    }
    \label{fig:cut_ty01_L20}
\end{figure}

\begin{figure*}[t]
    \centering
    \includegraphics[width=1\linewidth]{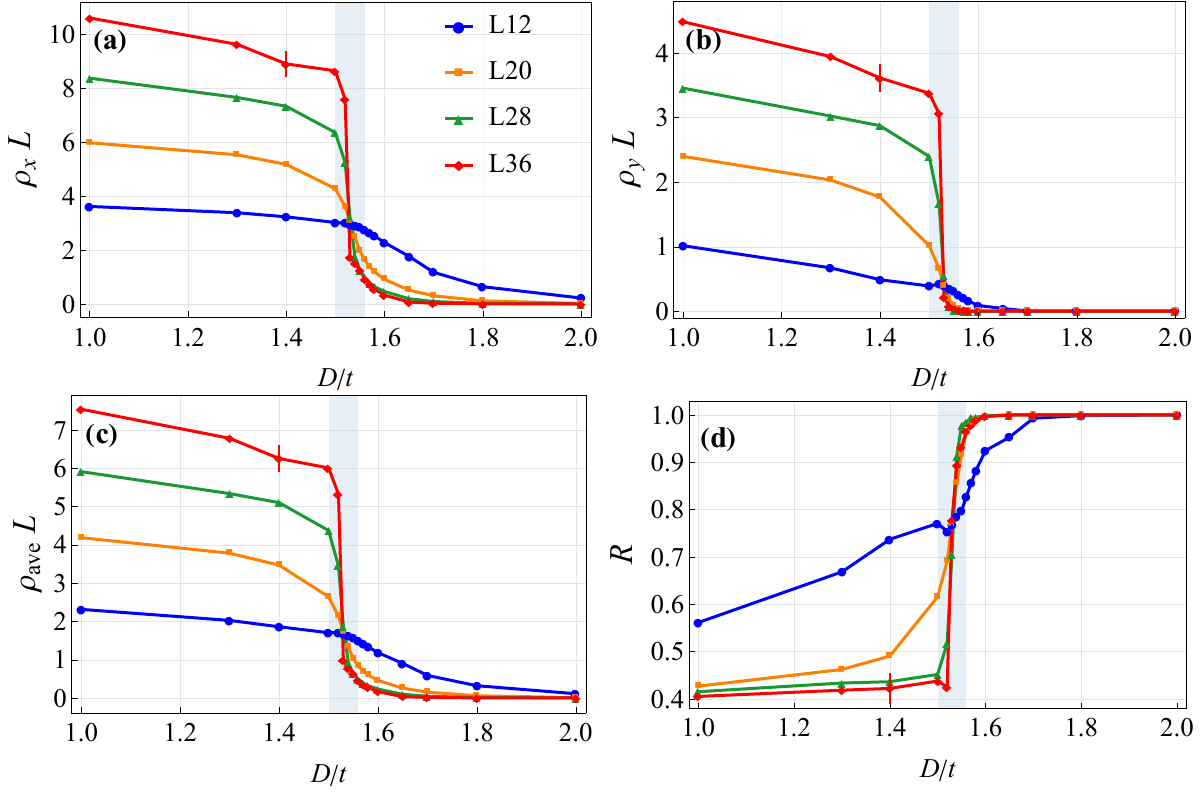}
    \caption{\textbf{Finite-size evolution of direction-resolved superfluid stiffness and transport anisotropy at fixed hopping anisotropy $t_y^{\rm eff}/t_x=0.1$.}
    Panels (a)--(c) show the scaled superfluid stiffnesses $\rho_{x}L$, $\rho_{y}L$, and $\rho_{\mathbf{ave}}L$, where $\rho_{\mathbf{ave}}=(\rho_{x}+\rho_{y})/2$, as functions of dipolar interaction strength $D/t$ for system sizes $L=12,20,28,36$ at half filling.
    Panel (d) shows the corresponding stiffness anisotropy
    $R=(\rho_{x}-\rho_{y})/(\rho_{x}+\rho_{y})$.
    As $D/t$ increases, $\rho_{y}$ is suppressed much earlier than $\rho_{x}$, leading to a rapid growth of $R$ toward unity and indicating strongly anisotropic transport near the SF--CB transition region.
    For the smaller system sizes, $\rho_{x}L$ and $\rho_{\mathbf{ave}}L$ display an apparent crossing near $D/t\approx1.54$, reminiscent of continuous-transition scaling.
    However, the larger-$L$ curves become substantially steeper rather than collapsing smoothly, indicating that the small-size crossing is better interpreted as a finite-size pseudo-critical feature than as definitive evidence for an asymptotically continuous transition.
    }
    \label{fig:ty01_rhoL_A_scaling}
\end{figure*}

\subsection{SF--CB evolution under strong anisotropy: continuous behavior on small lattices at $t_y^{\rm eff}/t_x=0.1$}
\label{subsec:cut_ty01}

We first examine the strongly anisotropic case $t_y^{\rm eff}/t_x=0.1$, which provides the clearest view of how Floquet-engineered kinetic anisotropy reshapes the competition between superfluid phase and checkerboard phase.
Figure~\ref{fig:cut_ty01_L20} shows the direction-resolved superfluid stiffnesses $\rho_{x}$ and $\rho_{y}$, the checkerboard structure factor $S(\pi,\pi)$, the compressibility $\kappa$, and the anisotropy indicator $R$ as functions of $D/t$ at half filling for $L=20$.

At weak dipolar interaction, the system is a \emph{compressible anisotropic superfluid}:
both directional stiffnesses are finite ($\rho_{x},\rho_{y}>0$), the compressibility is nonzero ($\kappa>0$), and $S(\pi,\pi)$ remains small.
Although the interaction is isotropic for perpendicular dipole polarization, the imposed hopping anisotropy already produces $\rho_{x}>\rho_{y}$ in the superfluid regime.

As $D/t$ increases, the superfluid response is progressively suppressed and becomes strongly direction-dependent.
The stiffness along the weak-hopping direction, $\rho_{y}$, drops rapidly and becomes nearly zero already within the finite-size transition interval, while $\rho_{x}$ remains finite over a broader range. This establishes a regime in which the residual phase-coherent transport is predominantly along the $x$ direction.

Concomitantly, the checkerboard structure factor $S(\pi,\pi)$ rises smoothly, signaling the smooth development of diagonal checkerboard correlations.
The same interval is accompanied by a pronounced collapse of the compressibility:
as shown in Fig.~\ref{fig:cut_ty01_L20}(b), $\kappa$ drops by several orders of magnitude and becomes consistent with zero on the large $D/t$ side.
Thus, the state is characterized not only by strong checkerboard correlations but also by \emph{incompressibility}, consistent with a checkerboard solid at half filling at larger $D/t$.

Figure~\ref{fig:cut_ty01_L20}(c) quantifies the transport anisotropy through $R$.
Starting from a moderate value on the superfluid side, $R$ rises steeply across the transition region and approaches $R\approx 1$ once $\rho_{y}$ is nearly quenched.
This shows that the loss of superfluid transport is not isotropic in practice: the weak-hopping direction loses stiffness first, while a residual $x$-direction response survives over a narrow finite-size interval.

As shown in Fig.~\ref{fig:cut_ty01_L20}, the evolution looks rather smooth for $L=20$.
Indeed, the directional stiffnesses decrease continuously, $S(\pi,\pi)$ grows smoothly and without an obvious macroscopic jump, and $\kappa$ is suppressed over a narrow yet finite interval.
From this small-system perspective, the SF--CB changeover is naturally resolved as an apparently continuous finite-size crossover region.

This small-size picture is refined below by explicit finite-size scaling.
As the system size increases, the same transition interval sharpens substantially, indicating that the apparently continuous behavior seen on the $L=20$ lattice is a finite-size manifestation of an underlying weakly first-order SF--CB transition.

\subsection{Finite-size evolution of anisotropic superfluid response and checkerboard ordering at $t_y^{\rm eff}/t_x=0.1$}
\label{subsec:ty01_fss_rho_S}

To clarify the nature of the SF--CB transition in the strongly anisotropic regime, we perform a combined finite-size analysis of both the superfluid response and the checkerboard structure factor at fixed hopping anisotropy $t_y^{\rm eff}/t_x=0.1$, tracking within the same narrow interaction interval how off-diagonal coherence collapses while diagonal checkerboard order develops as the system size increases.

Figure~\ref{fig:ty01_rhoL_A_scaling} presents the finite-size behavior of the direction-resolved superfluid stiffnesses for $L=12,20,28,36$, including $\rho_{x}L$, $\rho_{y}L$, and $\rho_{\mathbf{ave}}L$, together with the anisotropy ratio $R$.
Several robust trends emerge.

First, the superfluid response along the weak-hopping direction ($y$) is suppressed much more rapidly than along $x$ as $D/t$ increases.
This is already evident in the $L=20$ plots in Fig.~\ref{fig:cut_ty01_L20} and becomes even clearer in the finite-size results in Fig.~\ref{fig:ty01_rhoL_A_scaling}(b). Second, the anisotropy ratio $R$ rises sharply across the same interaction interval and approaches values close to unity, indicating that the remaining phase-coherent transport becomes nearly one-directional.
This behavior is consistent with a kinetically driven anisotropic loss of coherence: the weak-hopping direction loses stiffness first, while the $x$-direction response survives over a somewhat broader finite-size range. Third, the finite-size evolution of the scaled stiffnesses is nontrivial.
For the smaller system sizes, especially $L=12$, $20$, and $28$, $\rho_{x}L$ and $\rho_{\mathbf{ave}}L$ exhibit an apparent crossing near $D/t\approx1.54$, which at first sight resembles the standard scaling behavior expected near a continuous transition. The use of the scaled combination $\rho_s L$ follows from the standard quantum-critical finite-size form $\rho_s\sim L^{-(d+z-2)}$ in spatial dimension $d$. For a continuous transition in the present $(2+1)$-dimensional system with dynamical exponent $z=1$, one expects $\rho_s\sim L^{-1}$ at criticality, so $\rho_s L$ should be approximately size independent up to scaling corrections.
In the present result, however, the small-size crossing does not evolve into a clean common intersection as $L$ increases.
Instead, the larger-size curves ($L=36$) become significantly steeper within the same narrow interaction range.
A similar tendency is visible in $\rho_{y}L$, whose collapse also becomes increasingly abrupt for larger $L$.

\begin{figure}[t]
    \centering
    \includegraphics[width=\linewidth]{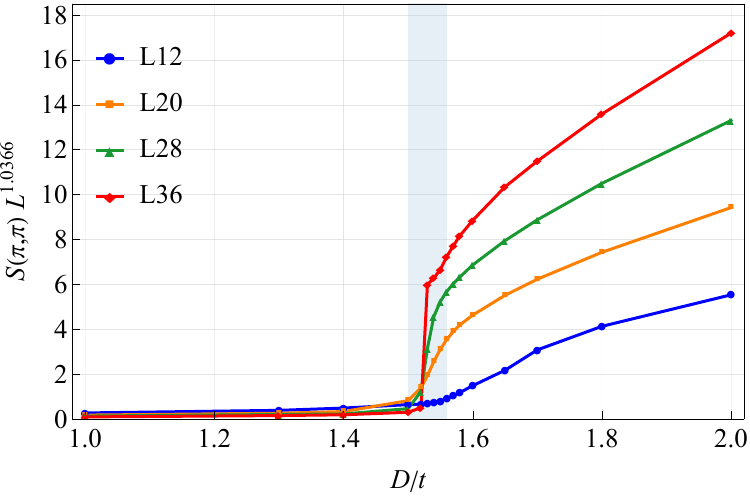}
    \caption{\textbf{Finite-size evolution of the checkerboard structure factor at hopping anisotropy $t_y^{\rm eff}/t_x=0.1$.}
    Shown is the checkerboard structure factor $S(\pi,\pi)$ for $L=12,20,28,36$, plotted in the scaled form $S(\pi,\pi) L^{1.0366}$ as a function of $D/t$.
    Combined with Fig.~\ref{fig:ty01_rhoL_A_scaling}, this behavior is consistent with a finite-size-rounded weakly first-order SF--CB transition.
    }
    \label{fig:ty01_strL_scaling}
\end{figure}

Complementary information is provided by the checkerboard structure-factor scaling in Fig.~\ref{fig:ty01_strL_scaling}, where we plot the scaled quantity $S(\pi,\pi) L^{1.0366}$ for the same set of system sizes.
Here the exponent $1.0366$ is chosen as the reference value $2\beta/\nu$ for the 3D Ising universality class, which serves as the natural continuous-transition benchmark for checkerboard ($Z_2$) ordering in a $(2+1)$-dimensional quantum lattice system.
We stress again that this is used only as a diagnostic scaling reference, not as evidence that the transition is asymptotically in the 3D Ising universality class.

The structure-factor results show the same qualitative finite-size evolution as the stiffness data: for $L=12$, $20$, and $28$ the scaled curves suggest an apparent crossing or near-crossing behavior, whereas the $L=36$ data sharpen much more abruptly in the same narrow interaction interval. This parallel evolution again supports a weakly first-order interpretation rather than a clean continuous-scaling scenario.

The most important observation is that the rapid increase of $S(\pi,\pi)$ occurs in essentially the same parameter region where the directional superfluid stiffnesses collapse most strongly, the anisotropy ratio $R$ rises sharply [Fig.~\ref{fig:ty01_rhoL_A_scaling}(d)], and the apparent $\rho_{x}L$ and $\rho_{\mathbf{ave}}L$ crossings are found for the smaller sizes.
This one-to-one correspondence indicates that the same narrow interaction interval controls both the loss of off-diagonal coherence and the establishment of checkerboard order.

From the finite-size perspective, the transition at $t_y^{\rm eff}/t_x=0.1$ is therefore best summarized as follows:
for small lattices, the result can mimic a continuous critical scenario;
as the system size increases, however, both the superfluid-collapse curves and the checkerboard-ordering curves sharpen markedly, supporting the interpretation of a weakly first-order SF--CB transition rounded by finite-size effects. Figs.~\ref{fig:cut_ty01_L20}, \ref{fig:ty01_rhoL_A_scaling}, and \ref{fig:ty01_strL_scaling} provide a consistent finite-size picture in which anisotropic superfluid transport is lost and checkerboard order is established within the same narrow transition interval.

\subsection{Checkerboard supersolidity upon doping away from half filling}
\label{subsec:V3_ss}

\begin{figure}[t]
    \centering
    \includegraphics[width=0.96\linewidth,trim=0cm 0cm 0cm 0cm,clip]{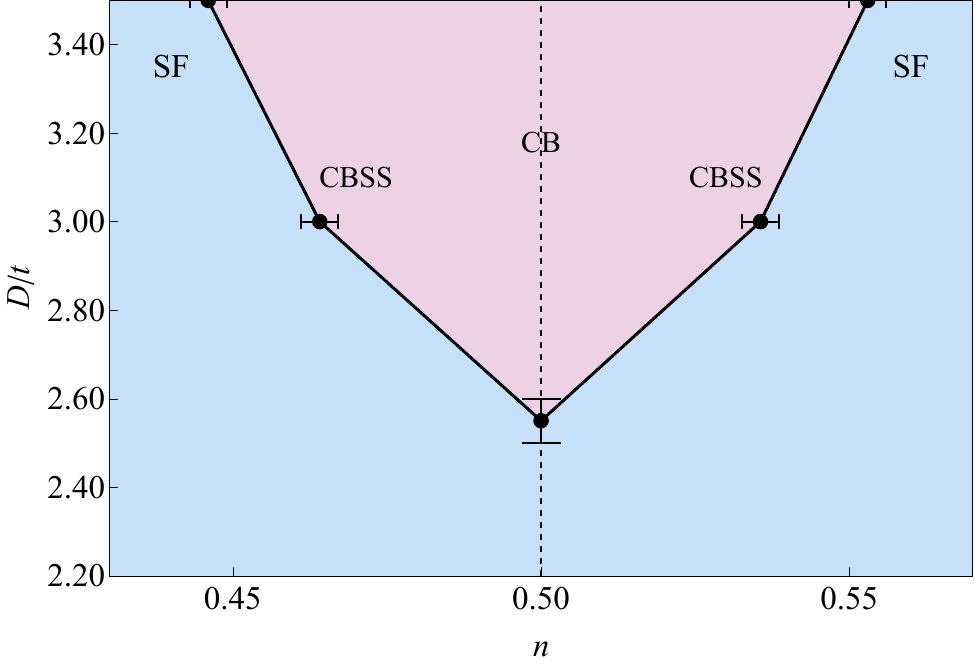}
    \caption{\textbf{Phase diagram in the $(n,D/t)$ plane summarizing the evolution of phase changes upon doping away from half filling at $t_y^{\rm eff}/t_x=0.5$.}
    The vertical line at $n=0.5$ corresponds to the commensurate checkerboard solid (CB).
    Upon doping, a narrow checkerboard supersolid (CBSS) lobe emerges on both sides of half filling, where checkerboard correlations remain substantial while superfluid stiffness becomes finite.
    Outside the lobe, the system is superfluid (SF).
}
    \label{fig:ty05_CBSS_phase_diagram}
\end{figure}

We next examine the phase diagram as a function of filling factor $n$ and interaction strength $D/t$. The results show that doping away from the half-filled checkerboard solid produces a regime with simultaneous density order and superfluidity. This behavior is summarized schematically in Fig.~\ref{fig:ty05_CBSS_phase_diagram} for $t_y^{\rm eff}/t_x=0.5$: the commensurate CB state is stabilized at $n=0.5$, while a narrow CBSS lobe appears upon doping on both sides. Outside this lobe, checkerboard correlations are strongly suppressed and the system enters SF regime.

\begin{figure}[t]
    \centering
    \includegraphics[width=\linewidth,trim=4.cm 0cm 5cm 0cm,clip]{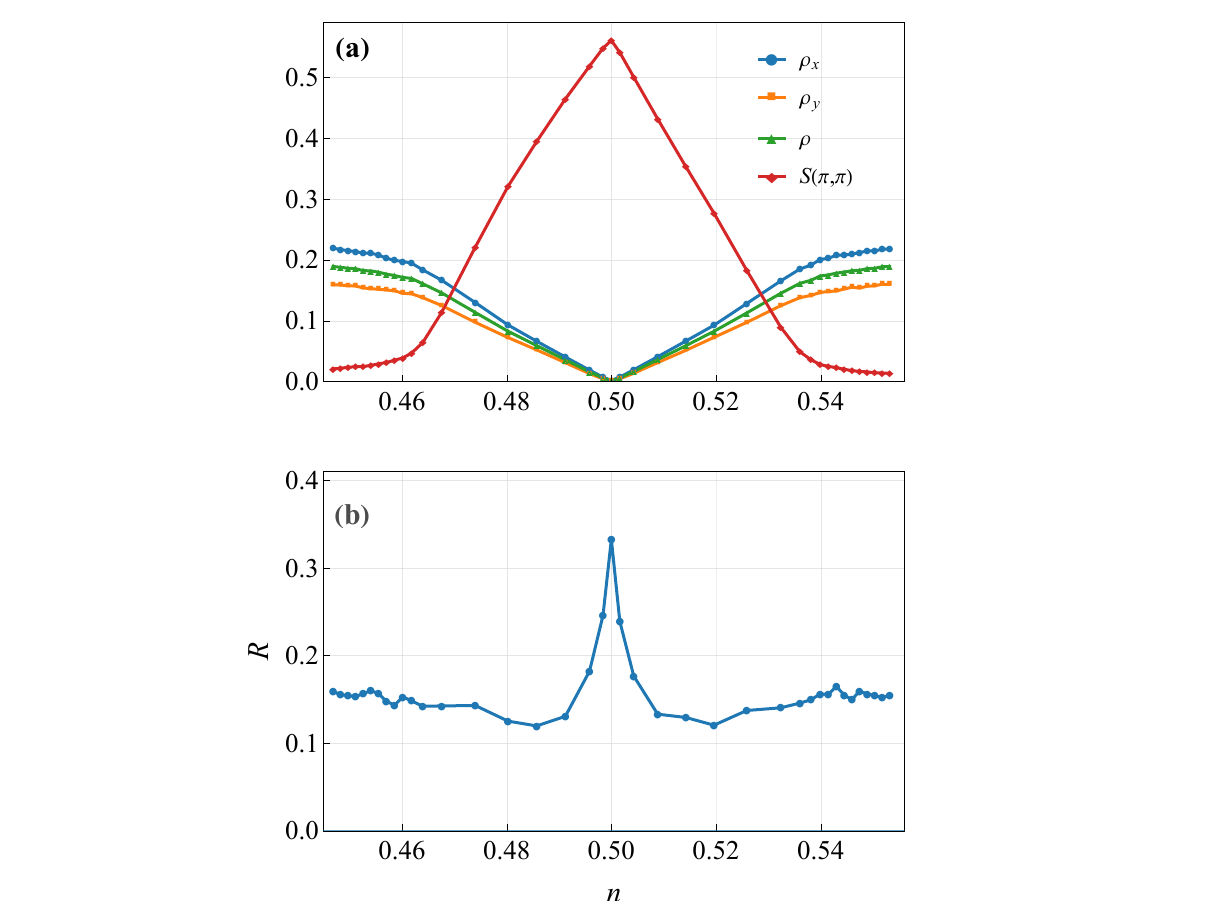}
    \caption{\textbf{Filling-resolved evolution of checkerboard order and superfluid response at $D/t=3.0$, $t_y^{\rm eff}/t_x=0.5$ for $L=20$.}
    As the filling approaches $n=0.5$, the checkerboard structure factor $S(\pi,\pi)$ is strongly enhanced, while the superfluid response is suppressed at the commensurate point and recovers upon doping away from half filling. On both sides of $n=0.5$, there exists a finite filling window where sizable checkerboard correlations coexist with nonzero superfluid stiffness, directly identifying a checkerboard supersolid regime.}
    \label{fig:V3_filling}
\end{figure}

Figure~\ref{fig:V3_filling} shows that the checkerboard structure factor $S(\pi,\pi)$ is maximized near $n=0.5$, signaling strong diagonal ordering. At the same time, the superfluid responses are strongly suppressed at half filling but remain finite immediately upon doping away from $n=0.5$. As a result, there is a finite interval on both sides of half filling in which appreciable checkerboard order coexists with nonzero superfluid stiffness, identifying a checkerboard supersolid.

\begin{figure}[t]
    \centering
    \includegraphics[width=\linewidth,trim=0cm 0cm 0cm 0cm,clip]{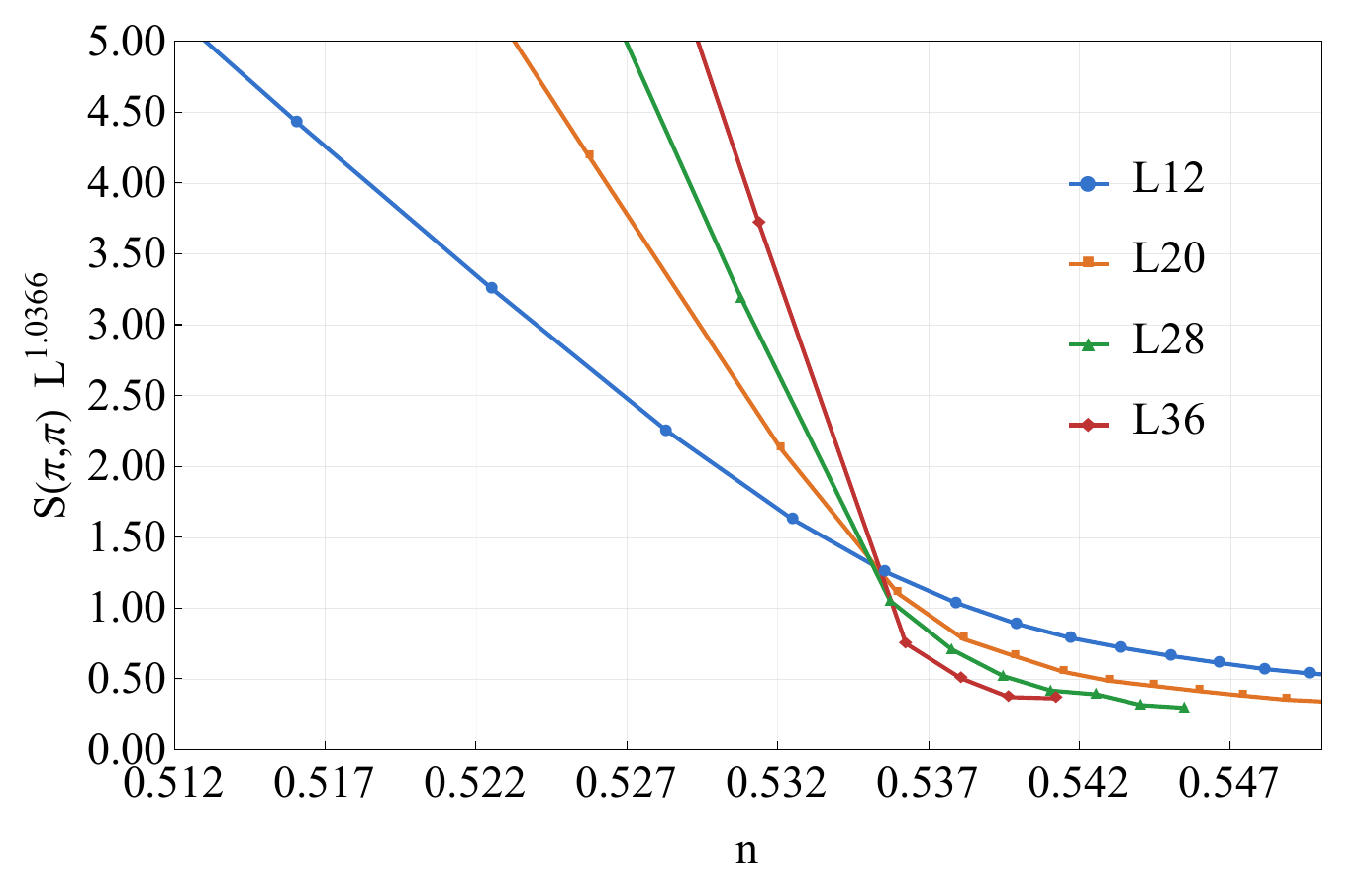}
    \caption{\textbf{Finite-size scaling of the checkerboard structure factor near half filling at $D/t=3.0$ and $t_y^{\rm eff}/t_x=0.5$.}
    The curves cross near $n=0.535\pm0.003$.
    This behavior supports the persistence of checkerboard ordering in the doped regime adjacent to $n=0.5$.}
    \label{fig:V3_scaling_S}
\end{figure}

The finite-size scaling of the checkerboard structure factor is shown in Fig.~\ref{fig:V3_scaling_S}. 
Here the rescaled quantity is chosen as $S(\pi,\pi) L^{1.0366}$, where the exponent $1.0366$ is the reference value of $2\beta/\nu$ for the three-dimensional Ising universality class.
This is the natural scaling benchmark for checkerboard ($Z_2$) ordering in a $(2+1)$-dimensional quantum system. As shown in Fig.~\ref{fig:V3_scaling_S}, the rescaled structure-factor curves cross near $n=0.535\pm0.003$, which is compatible with a CBSS--SF transition governed by 3D Ising scaling of the checkerboard order parameter.

\begin{figure*}[t]
    \centering
    \includegraphics[width=\linewidth,trim=0cm 0cm 0cm 0cm,clip]{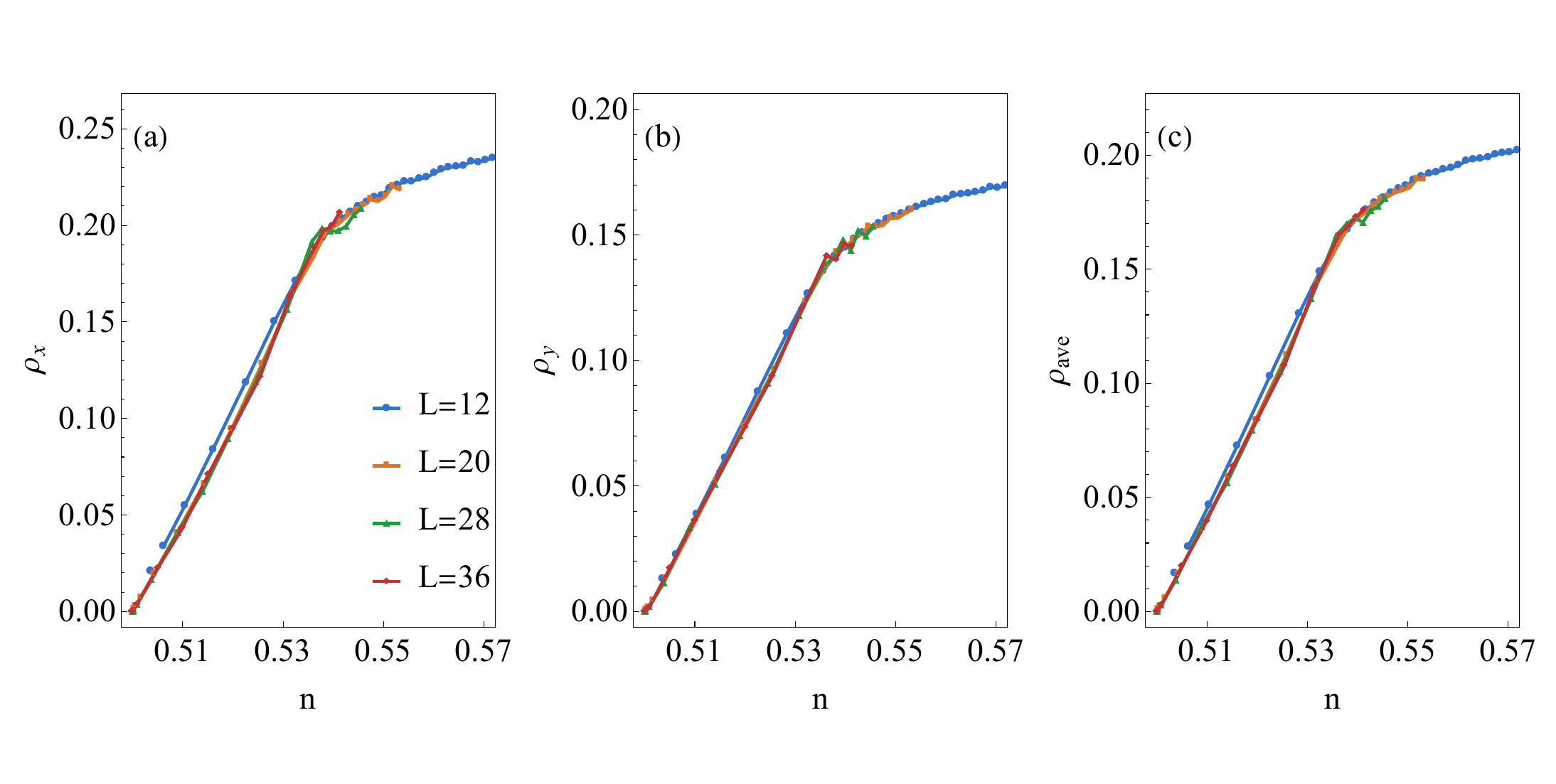}
    \caption{\textbf{Finite-size behavior of the superfluid stiffness near half filling at $D/t=3.0$ and $t_y^{\rm eff}/t_x=0.5$.}
    In the same filling window where checkerboard correlations remain substantial, the superfluid stiffness stays finite for all system sizes studied.
    The stiffness curves for different $L$ nearly overlap, indicating weak finite-size effects rather than a suppression of superfluidity with increasing system size.
    Such behavior is consistent with a stable nonzero superfluid response in the doped regime adjacent to $n=0.5$.}
    \label{fig:V3_scaling_rho}
\end{figure*}

The superfluid sector exhibits another important finite-size behavior.
As shown in Fig.~\ref{fig:V3_scaling_rho}, the superfluid stiffness remains finite throughout the doped region and, crucially, does not decrease as the system size increases.
Instead, the result for different lattice sizes nearly overlap within numerical uncertainty.
Here the near-collapse of the stiffness data is more naturally interpreted as evidence of weak finite-size effects, indicating that the superfluid response remains stable in the thermodynamic limit.
This finite-size phenomenology is qualitatively compatible with earlier isotropic square-lattice studies in which doped checkerboard supersolid regimes were identified from the coexistence of robust density order and finite superfluid response\cite{Ohgoe2012,Capogrosso2010}.

Thus, Figs.~\ref{fig:ty05_CBSS_phase_diagram}--\ref{fig:V3_scaling_rho} identify a checkerboard supersolid on both sides of half filling at $t_y^{\rm eff}/t_x=0.5$: checkerboard density correlations remain substantial upon doping away from the commensurate solid, while the superfluid stiffness becomes finite and remains essentially size-independent in the same filling window.

\section{Conclusion}
\label{sec:conclusion}

We have studied the interplay between Floquet-engineered hopping anisotropy and dipolar interactions in the square-lattice hard-core Bose--Hubbard model, with particular focus on the competition between superfluid coherence and checkerboard density ordering.
By combining direction-resolved superfluid diagnostics, compressibility, and structure-factor measurements over a broad parameter range, we showed that tuning the effective hopping ratio $t_y^{\rm eff}/t_x$ does not merely rescale the kinetic energy, but qualitatively reshapes both the location of the SF--CB transition.

Our first main result is the half-filling phase diagram in the $(t_y^{\rm eff}/t_x,D/t)$ plane.
As the hopping anisotropy becomes stronger, the interaction strength required to stabilize checkerboard order is reduced.
This demonstrates that suppressing motion along one spatial direction strongly enhances interaction-driven crystallization, even though the dipolar interaction itself remains isotropic.
Thus, Floquet-controlled kinetic anisotropy provides an efficient route for shifting the balance from off-diagonal coherence toward diagonal density order.

Our second main result concerns the nature of the half-filling SF--CB transition in the strongly anisotropic regime.
From finite-size scans and scaling analyses at $t_y^{\rm eff}/t_x=0.1$, we found that the transition is resolved on finite systems as a narrow interval over which the directional superfluid stiffnesses collapse rapidly, the anisotropy ratio grows strongly, the compressibility is suppressed, and the checkerboard structure factor rises sharply.
For smaller system sizes, this interval can resemble a smooth crossover and may mimic aspects of continuous critical behavior.
However, as the system size increases, the evolution becomes progressively sharper.
These results indicate a narrow, finite-size-rounded weakly first-order SF--CB transition, although a fully definitive thermodynamic diagnosis would still require larger systems and additional first-order probes.

A third important result emerges away from half filling.
On the doped sides of the checkerboard plateau, the incompressible CB solid gives way to regimes in which checkerboard correlations remain sizable while the superfluid response becomes finite again.
This coexistence of diagonal order and off-diagonal coherence identifies a checkerboard supersolid near commensurate filling.

Overall, our results establish a compact and experimentally relevant picture of anisotropy-controlled phase competition in hard-core dipolar lattice bosons.
At half filling, hopping anisotropy shifts the SF--CB transition to lower interaction strength and produces a narrow transition regime with weakly first-order transition.
Away from half filling, the same anisotropic setting yields narrow checkerboard-supersolid windows adjacent to the checkerboard plateau.
These findings provide a useful framework for analyzing Floquet-engineered optical-lattice realizations, where directional tunneling can be tuned and both density order and transport anisotropy can be probed independently.

\begin{acknowledgments}
C. Zhang would like to thank Lode Pollet for helpful discussion. 
\end{acknowledgments}

\bibliography{anisotropic_refs}

\end{document}